\documentclass[conference]{IEEEtran}
\usepackage{url}
\usepackage{graphicx}
\usepackage{algorithmic}
\usepackage{algorithm}
\usepackage{amsmath}
\usepackage{amsfonts}
\usepackage{color}
\usepackage[normalem]{ulem}
\usepackage{dsfont}
\usepackage{listings}
\usepackage[utf8]{inputenc}
\graphicspath{ {./images/} }
\usepackage{acronym}
\usepackage{cite}
\usepackage{array}
\usepackage{subfigure}
\usepackage{dblfloatfix}
\usepackage{soul}
\def\BibTeX{{\rm B\kern-.05em{\sc i\kern-.025em b}\kern-.08em
    T\kern-.1667em\lower.7ex\hbox{E}\kern-.125emX}}
    \newcolumntype{P}[1]{>{\centering\arraybackslash}p{#1}}
\newcolumntype{M}[1]{>{\centering\arraybackslash}m{#1}}
\title{Analysis of Climatic Trends and Variability in Indian Topography}

\author{ Ayush Prusty$^{[1]}$, Akshita Gupta$^{[2]}$, and Vivek Ashok Bohara$^{[2]}$\\
$^{[1]}$Department of Computer Science \& Social Science Engineering,\\ $^{[2]}$Department of Electronics \& Communication Engineering,\\
Indraprastha Institute of Information Technology Delhi (IIITD), New Delhi 110020, India.\\
Email: ayush20427@iiitd.ac.in, akshitag@iiitd.ac.in, and vivek.b@iiitd.ac.in
}

\begin{document}

\maketitle
\thispagestyle{empty}
\pagestyle{empty}
\begin{abstract}
The climatic change is one of the serious concerns nowadays.  The impacts of climate change are global in scope and unprecedented in scale. Moreover, a small perturbation in climatic changes affects not only the pristine ecosystem but also the socioeconomic sectors. Specifically, the affect of climatic changes is related to frequent casualties. This makes it essential to dwelve deeper into analyzing the socio-climatic trends and variability. This work provides a comprehensive analysis of India's climatic trends, emphasizing on regional variations and specifically delving into the unique climate of Delhi. Specifically, this research unveils the temporal and spatial variations in temperature patterns by amalgamating extensive datasets encompassing India's diverse landscapes. The study uses advanced statistical tools and methodologies to scrutinize temperature's annual and seasonal variability. The insights drawn from this rigorous analysis may offer invaluable contributions to regional planning strategies, adaptive measures, and informed decision-making amidst the complex impacts of climate change. By bridging the gap between broader climatic trends and localized impacts, this research aims to facilitate more effective measures to mitigate and adapt to the multifaceted challenges of climate change, ensuring a more nuanced and tailored approaches. We utilized the Mann-Kendall test and Theil-Sen's slope estimator to analyze the trends and variability of the climatic conditions over the decades. The results demonstrate that temperature variations have increased over 0.58$^{o}C$ on average over the last decade. Moreover, over last decade the variability of Indian states shows that Lakshadweep faced the highest change (0.87$^{o}C$), highlighting coastal vulnerability, while Tripura observed the least change of 0.07$^{o}C$.
\end{abstract}
\textbf{Abstract}

The climatic change is one of the serious concerns nowadays.  The impacts of climate change are global in scope and unprecedented in scale. Moreover, a small perturbation in climatic changes affects not only the pristine ecosystem but also the socioeconomic sectors. Specifically, the affect of climatic changes is related to frequent casualties. This makes it essential to dwelve deeper into analyzing the socio-climatic trends and variability. This work provides a comprehensive analysis of India's climatic trends, emphasizing on regional variations and specifically delving into the unique climate of Delhi. Specifically, this research unveils the temporal and spatial variations in temperature patterns by amalgamating extensive datasets encompassing India's diverse landscapes. The study uses advanced statistical tools and methodologies to scrutinize temperature's annual and seasonal variability. The insights drawn from this rigorous analysis may offer invaluable contributions to regional planning strategies, adaptive measures, and informed decision-making amidst the complex impacts of climate change. By bridging the gap between broader climatic trends and localized impacts, this research aims to facilitate more effective measures to mitigate and adapt to the multifaceted challenges of climate change, ensuring a more nuanced and tailored approaches. We utilized the Mann-Kendall test and Theil-Sen's slope estimator to analyze the trends and variability of the climatic conditions over the decades. The results demonstrate that temperature variations have increased over 0.58$^{o}C$ on average over the last decade. Moreover, over last decade the variability of Indian states shows that Lakshadweep faced the highest change (0.87$^{o}C$), highlighting coastal vulnerability, while Tripura observed the least change of 0.07$^{o}C$.

\section{Introduction}
Climate change is an omnipresent concern on a global scale and manifests in diverse regional nuances across different geographical landscapes. The impacts of climate change span across temperature shifts, altered precipitation patterns, and ecological transformations, delineating the complex tapestry of environmental change\cite{ref2}. Although global trends have been captured comprehensively, the regional variations demand nuanced exploration for informed decision-making and adaptive strategies. The recent climatic changes witnessed all over the globe are a potential threat to humans today. Significant changes have been observed in all the important climate variables. The rise in global temperature is among the fundamental changes catalyzing the extremities in a large number of climatic factors. For instance,~\cite{climate_change_1} reported a rise of 0.40$^{o}C$ in the global surface during the last century. 

India has a heterogeneous tapestry of landscapes and has witnessed consistent shifts in its climatic patterns over recent decades. According to~\cite{India1}, India was ranked fourth among the list of countries most affected by climate change in 2015. Calculations in 2021 showed that India should increase its climate commitments by 55\% to give the world a 50\% chance of avoiding a temperature rise of 2 degrees or more~\cite{India2}. Raising temperatures on the Tibetan Plateau are forcing the Himalayan glaciers to recede, endangering the flow of important rivers such as the Yamuna, Brahmaputra, and Ganges. Climate change is causing an increase in the frequency and intensity of heat waves in India. It is anticipated that states like Assam will see an increase in the frequency of severe landslides and floods~\cite{India4}. India's climate change performance score is ranked eighth out of 63 countries that would produce 92\% of the world's greenhouse gas emissions in recent years~\cite{India5}.

 Numerous studies have elucidated the observed climate variability, changes in precipitation, and temperature trends across the country \cite{dash2009changes,singh2015trend,shekhar2010climate}. For instance, in~\cite{dash2009changes}, the authors examined possible changes in the frequency of rain events in India in terms of their duration and intensity per day. Similarly, the authors in~\cite{singh2015trend} studied spatial and temporal variability in temperature over selected stations of the Sutlej basin located in the North-Western Himalayan region in India. Further, the authors in~\cite{shekhar2010climate,dad2021time} described the trends in weather and climate over the western Himalayas. Although recent research on the impact of climate change has been targeting specific locations in India; however, we intend to produce a study that generates a comparison of trends all across India with a special focus on the climatic change of Delhi. Nevertheless, uncertainties persist regarding the regional impacts of this climatic evolution, especially considering India's diverse topography and varied climatic zones \cite{ref3}. This research holistically explores India's climatic trends, amalgamating comprehensive data encompassing the nation's landscape. %Additionally, it depends on the specific climate data pertinent to Delhi, the bustling metropolis, and the nation's capital. 
\begin{figure*}[t]
  \centering
  \includegraphics[width=\textwidth]{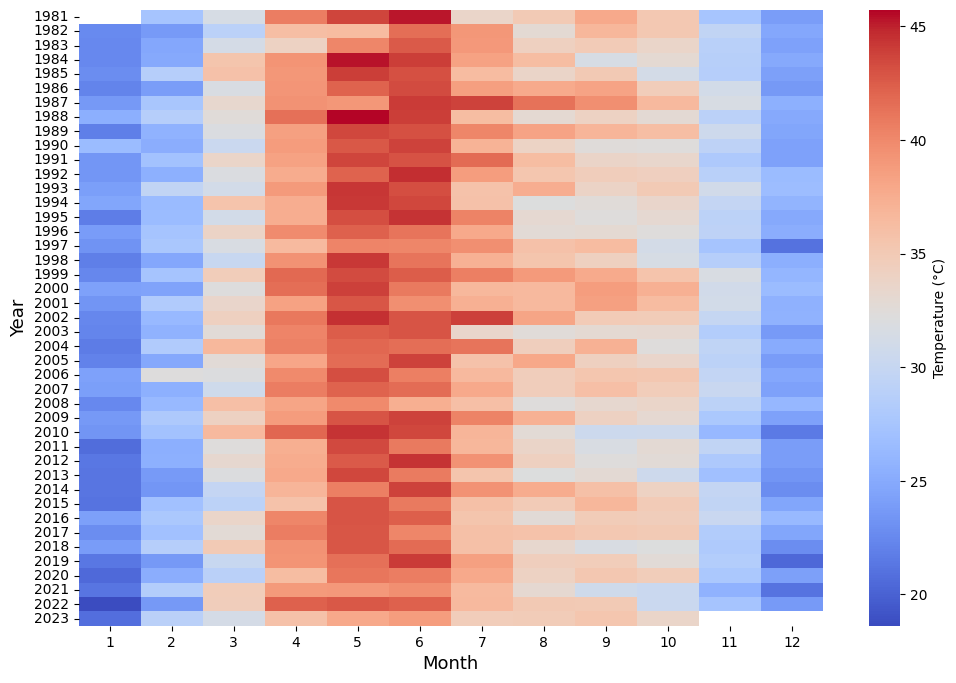}
  \caption{Daily average temperature in New-Delhi ($^o C$)}
  \label{fig:delhi1}
\end{figure*}
 Including India's diverse terrains and the specific focus on Delhi enables a comprehensive examination of varied climate patterns, offering valuable insights for regional planning, sustainable development, and climate adaptation strategies. This research aspires to unravel the temporal and spatial variations in India's climate as well as the unique climatic trends within Delhi by employing statistical tools and methodologies.
A meticulous examination of India's climate landscape and the exclusive analysis of Delhi's climate patterns aim to provide essential data-driven insights for policymakers, environmentalists, and local communities. These insights will play a crucial role in guiding informed decisions, formulating adaptive measures, and bolstering resilience against the multifaceted impacts of climate change at both national and local scales. The significance of this research lies in comprehending the intricate fabric of regional climatic changes and in the practical application of these insights. Understanding local-scale climate trends is indispensable for sustainable development, resource allocation, and formulating region-specific climate adaptation strategies. By unraveling the temporal variations and trends in temperature within Delhi, this study aspires to bridge the gap between global climate models and localized climate dynamics. Some of the major contributions of this work are summarized below:
\begin{enumerate}
   \item We utilize Indian climatic data from the years 1980 to 2020 to scrutinize annual and seasonal temperature variability and patterns across India. We configured temperature trends across the country based on the variability and classified the different Indian states and union territories based on the temperature variability. 
   \item Specifically, we focus on Delhi and show the temperature variation trends in recent years. We utilize the data from the years 2016 to 2020 to analyze the seasonal trends across the city.
   \item We utilize various statistical tools, including parametric and non-parametric tests such as the Mann–Kendall test and Theil-Sen's slope estimator, to dissect the intricacies of Delhi's climatic changes.
   \item The test results are compiled to draw insights into the nature of temperature change in India. 
\end{enumerate}
%Through this comprehensive analysis, the study endeavors to contribute to the broader understanding of regional climate variability, providing crucial data-driven insights that can aid policymakers, environmentalists, and local communities in making informed decisions and formulating adaptive measures to mitigate the impacts of climate change. 
By examining India's data, and then focusing on Delhi's climate time series, we can unravel trends, anomalies, and fluctuations influenced by natural and anthropogenic factors. The insights from such analyses can influence policies, urban planning, and adaptation strategies, contributing to building climate resilience and sustainable development for Delhi and offering valuable lessons for other urban centers grappling with similar global challenges. 
The rest of the paper is organized as follows: Section II delves into the detailed explanation of data utilized and processes used to discover variability and trends. In Section III, we analyze the results and discuss the inferences. Section IV concludes the paper and discusses the future scope.

\section{Materials and methods}
In this section, we discuss the processing and preparation of the data set and the statistical tools utilized for observing socio-climatic trends across India.

% India lies between latitudes 8° 4'N and 37° 6'N, and the longitudes 68°7'E and 97° 25'E. Delhi, the bustling capital of India, is nestled between 28.40°N latitude and 77.13°E longitude, situated within the larger Indo-Gangetic Plain. Spanning an area of approximately 1,484 square kilometers, Delhi encompasses a diverse landscape.

\subsection{Study Area}
India lies in the Northern Hemisphere. It serves as a crucible of historical, political, and cultural significance. India primarily experiences tropical weather. The northern region enjoys a dry environment similar to that of the equatorial regions throughout the summer because of the sun's position\cite{climate2}. India's unique location in relation to the Asian continent and the Indian Ocean results in a tropical monsoon climate. The summers are scorching, and the winters are mildly cold on the Indian subcontinent. Delhi, the bustling capital of India, encompasses a diverse landscape. The city's climate embodies a distinctive amalgamation of subtropical and semi-arid influences characterized by extreme variations in temperature and seasonal dynamics. According to \cite{krishnanimpact}, a significant portion of Delhi's climate lies under the Köppen climate classification of Cwa (i.e., monsoon-influenced humid subtropical climate, where the average temperature for the coldest month is above 0°C, at least one month's average temperature above 22°C, and at least four months averaging above 10°C). Fig.~\ref{fig:delhi1} presents Delhi's climate pattern over a period of 20 years, i.e., 1996-2016. Delhi's weather can be partitioned into distinct seasons: a scorching summer (April–June) marked by sweltering temperatures, often soaring above 40°C (104°F); a monsoon season (July–September) bringing relief in the form of sporadic yet heavy rainfall; a transitional post monsoon (October–November) with diminishing rainfall and dropping temperatures; and a relatively chilly winter (December–February) with temperatures occasionally dropping to 3-4°C (37-39°F) and foggy conditions, particularly in January. Delhi's unique urban setting amid rapid urbanization and population growth makes it an intriguing choice to analyze its temperature trends. The city's climatic dynamics, impacted by its urban sprawl, industrial activities, vehicular pollution, and human-induced factors, present an interesting and intricate matrix for climate studies. Moreover, the urban heat island effect, prevalent in metropolitan regions like Delhi due to dense habitation and concrete structures, adds a layer of complexity to its climatic evolution.
 Juxtaposed with historical weather records, Delhi's urban landscape offers an excellent opportunity to explore the interplay between urbanization, human activities, and climatic variations. Analyzing Delhi's climate time series aids in understanding the impact of urbanization on the local climate, facilitating the formulation of sustainable urban development strategies and climate-resilient infrastructure.

\subsection{Data Set}
For the analysis conducted in this study, we have collected data from two primary sources of climate data, namely a) the Climate Change Knowledge Portal (CCKP)~\cite{data1} and b) the National Renewable Energy Laboratory (NREL)~\cite{data2}. The Indian data has been taken from CCKP and it contains data from 1901-2020. the features of that dataset are state, period, and subsequent years. The data for Delhi has been taken from NREL, and the timeline for its data is 2016-2020. The two data sources are described in detail as follows:
\subsubsection{Climate Change Knowledge Portal (CCKP)} The CCKP provided comprehensive historical climate data for India~\cite{data1}. This included temperature, precipitation, sea-level rise datasets, and various climate-related vulnerabilities and impacts. The data spanned a significant timeline, enabling a detailed examination of long-term climate patterns and changes across the Indian subcontinent.
  \subsubsection{National Renewable Energy Laboratory (NREL)} Specific climate data about Delhi, the focus region of the proposed work is obtained from the NREL database~\cite{data2}. This dataset encompassed granular and localized climate information relevant to Delhi, offering detailed insights into the temperature variations, precipitation patterns, and other meteorological parameters specific to the region.
  
\subsection{Preprocessing}
The acquired datasets were analyzed and processed to comprehensively assess climate trends and variations. Raw datasets from CCKP and NREL were subjected to meticulous preprocessing steps. The data was cleaned, including removing duplicates and replacing null fields. Time series analysis techniques were employed to explore and identify temporal patterns, trends, and variations in temperature and precipitation datasets. This involved employing statistical methods and visualization tools to depict long-term changes and seasonality within the climate data. Various statistical techniques were employed to quantify and evaluate the observed trends and variability in the climate data.

% \begin{enumerate}
% \renewcommand{\labelenumi}{\alph{enumi})}

%     \item Mann-Kendall Test: The Mann-Kendall test detected monotonic trends in the time series data. This non-parametric test helped identify upward or downward trends in temperature and precipitation over the study periods.
    
%     \item Sen's Slope Estimator: Sen's slope estimator provided a robust method for estimating the magnitude and direction of trends in the climate variables. It facilitated a more nuanced understanding of the rate of change in temperature and precipitation data.
% \end{enumerate}

\subsection{Testing}
Various statistical methods, including parametric and non-parametric, have been applied to detect trends and other changes in climatic variables. However, non-parametric methods are usually considered inadequate. This is mostly because these tests do not assume a normal distribution for variables of interest and are robust to the influence of extremes \cite{articleNN}. Specifically, these techniques are deemed highly preferential because untransformed climatic data are often distinctly non-normal\cite{chattopadhyay2016long}. For the present study, we used the non-parametric Mann–Kendall test \cite{mann1945nonparametric, kendall1948rank} to assess the significance of trends in precipitation and temperature data on monthly, seasonal, and annual scales. The authors in~\cite{MKandTS} utilized Theil-Sen median trend analysis, and the Mann-Kendall method was used to analyze the spatiotemporal variation of vegetation coverage in the Belt and Road Initiative (BRI) region from 1982 to 2015. The null hypothesis in the test is that there is no significant trend within the time series. When rejected, this hypothesis indicates a trend, which can be either positive or negative, as described by its score.

\vspace{\baselineskip}
\subsubsection{Mann-Kendall Test}

The Mann–Kendall statistic $S$ of the series $x$ is given by~\cite{MKandTS}:
\begin{equation}
    S = \sum_{i=1}^{n-1} \sum_{j=i+1}^{n} \text{sgn}(x_j - x_i),
\end{equation}
where $x_i$ and $x_j$ are the annual values in years $i$ and $j$ ($j > i$) respectively. $n$ is the number of observations and $\text{sgn}$ is the signum function~\cite{data3}. %Each data point $x_i$ is compared with the rest of the data points $x_j$ using the expression:
%\begin{equation}
 %   \text{sgn}(x_j - x_i) = 
 %   \begin{cases}
 %       +1 & \text{if } (x_j - x_i) > 0, \\
%        0 & \text{if } (x_j - x_i) = 0, \\
%        -1 & \text{if } (x_j - x_i) < 0.
%    \end{cases}
%\end{equation}

The variance associated with $S$ is calculated as~\cite{MKandTS}:
\begin{equation}
    \text{Var}(S) = n(n - 1)(2n + 5) - \sum_{k=1}^{m} t_k(t_k - 1)(2t_k + 5) / 18,
\end{equation}
where $m$ is the number of tied groups or ties of the sample time series and $t_k$ is the number of data points in group $k$.

For $n > 10$, the test statistic $Z(S)$ is calculated as~\cite{mann1945nonparametric}:
\begin{equation}
    Z(S) = 
    \begin{cases}
        \frac{S - \bar{S}}{\sqrt{\text{Var}(S)}} & \text{if } S > 0, \\
        0 & \text{if } S = 0, \\
        \frac{S + 1}{\sqrt{\text{Var}(S)}} & \text{if } S < 0,
    \end{cases}
\end{equation}
where $\bar{S}$ is the mean of statistic $S$. The positive values of $Z(S)$ indicate increasing trends, while negative values suggest decreasing trends. For statistical significance at $\alpha = 0.05$, if $|Z(S)| > Z_{1-\alpha/2}$, it is treated as significant.
\begin{figure*}[t]
\centering
  \includegraphics[width=\linewidth, trim={0 0 0 1cm},clip]{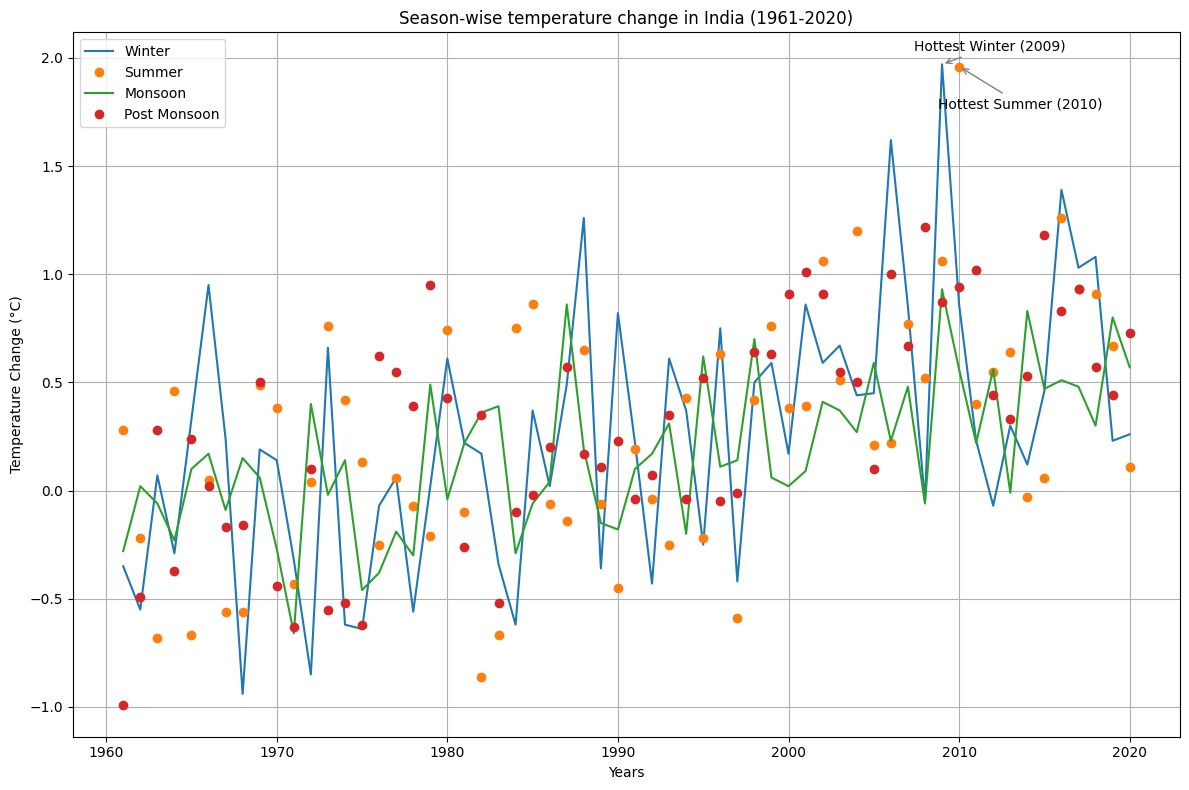}
  \caption{Season-wise average temperature change in India}
  \label{fig:wide_image}
\end{figure*}
%\begin{figure}
 %   \centering
 %   \includegraphics[width=\linewidth]{download_3.png}
  %  \caption{Caption}
 %   \label{fig:enter-label}
%\end{figure}
\subsubsection{Theil-Sen Approach}
To quantify significant linear trends, the Theil-Sen Approach (TSA) is used. The slope $Q$ between any two values of a time series data $x$ is estimated as~\cite{MKandTS}:
\begin{equation}
    Q = \frac{x_k - x_j}{k - j}, k \neq j.
\end{equation}

The overall estimator of the slope $Q$ by Sen’s method is given as~\cite{MKandTS}:
\begin{equation}
    Q^* = 
    \begin{cases}
        Q\left(\frac{N + 1}{2}\right), & \text{if } N \text{ is odd,} \\
        \frac{1}{2}\left(Q\left(\frac{N}{2}\right) + Q\left(\frac{N}{2} + 1\right)\right), & \text{if } N \text{ is even,}
    \end{cases}
\end{equation}
where $N = \frac{n(n-1)}{2}$ possible values of $Q$ exist. 
For calculating 95\% confidence intervals using non-parametric technique, $C_\alpha$ is calculated as~\cite{MKandTS}:
\begin{equation}
    C_\alpha = Z_{1 - \alpha/2} \sqrt{\text{Var}(S)}.
\end{equation}
The confidence limits are defined by the $M_1^{th}$ and $(M_2 + 1)^{th}$ largest of the ordered estimates of $Q$, with interpolation as appropriate for non-integer values of $M_1$ and $M_2$. The indices $M_1$ and $M_2$ are determined from~\cite{MKandTS}:
\begin{align}
    M_1 &= N - \frac{C_\alpha}{2}, \\
    M_2 &= N + \frac{C_\alpha}{2}.
\end{align}
Standardized anomalies of annual precipitation are calculated as follows:
\begin{equation}
    Z = \frac{X_i - X}{s},
\end{equation}
where $Z$ represents standardized precipitation; $X_i$ is the annual precipitation.% $T_{\text{max}}$, and $T_{\text{min}}$ are the maximum and minimum temperature of a particular year; $X$ is the long-term mean annual precipitation, $T_{\text{max}}$, and $T_{\text{min}}$ for observation and $s$ is the standard deviation of annual precipitation. In our paper, we have considered the observations from 1980-2017 to calculate $T_{\text{max}}$ and $T_{\text{min}}$.

Integrating these diverse datasets and analytical methods allows a comprehensive examination of climate trends, enabling a deeper understanding of India's temporal and spatial variability, specifically Delhi's climate.

\begin{figure*}
     \centering
  \includegraphics[width=\linewidth]{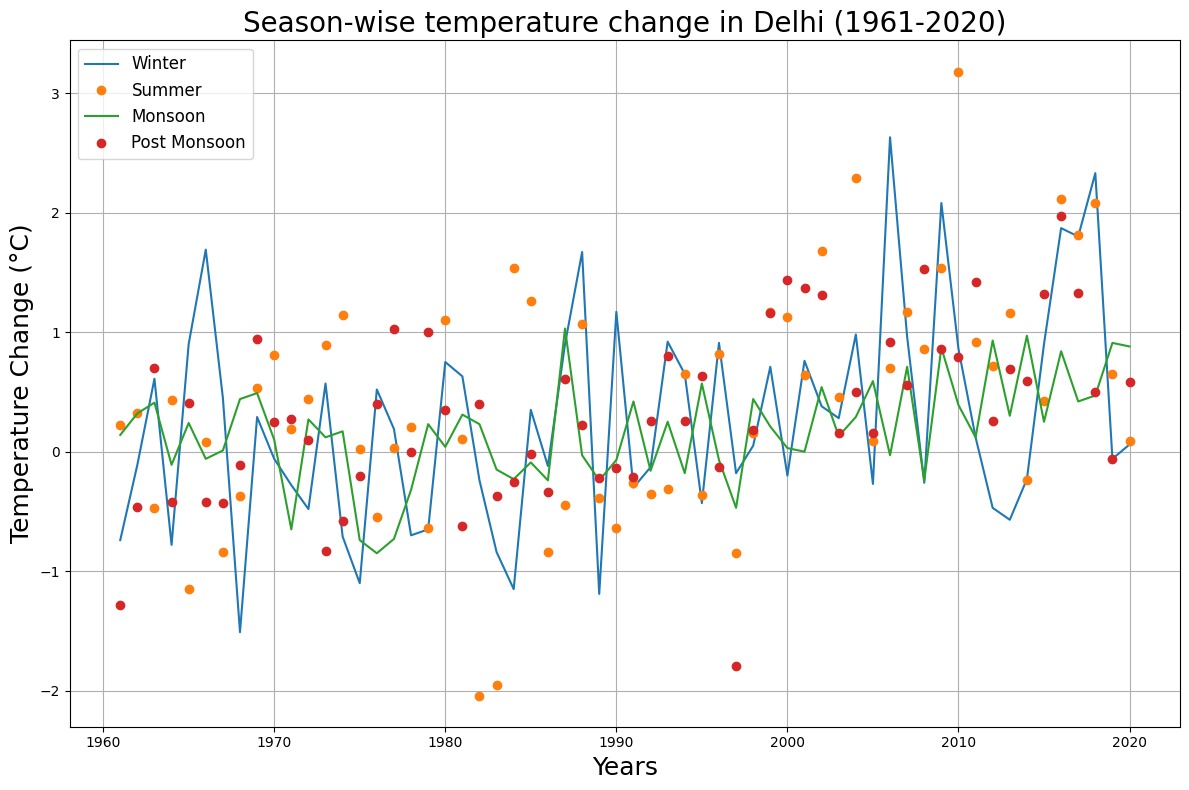}
  \caption{Seasonal trends for Delhi across the years}
  \label{Seasonal_Trends_across_the_years}
\end{figure*}

\section{Results}
This section shows the performance of the trend and variability tests on the considered data set. Utilizing the statistical tools in Section 2.D, we classified the different Indian states and union territories (UTs) based on temperature variability. Specifically, we focus on Delhi and show the temperature variation trends in the recent years\footnote{The study is agnostic to the region considered in the paper. A similar trend can be analyzed for other regions depending on the data availability for the regions.}. 
\begin{table}[h]
    \centering
     \caption{Temperature variability for Indian states and UTs}
    \begin{tabular}{|c|M{3 cm}|}\hline
         State/UTs& Mean temperature change over last decade~$(^{o}C)$\\\hline
        Lakshadweep & $0.87$\\\hline
        Andaman and
Nicobar Islands & $0.79$\\\hline
Tamil Nadu& $0.79$\\\hline
Puducherry& $0.78$\\\hline
Himachal Pradesh& $0.80$\\\hline 
Uttarakhand& $0.79$\\\hline 
Delhi& $0.76$\\\hline
Chandigarh& $0.75$\\\hline
Haryana&$0.70$\\\hline
Tripura& $0.07$\\\hline 
Mizoram& $0.11$\\\hline
Meghalaya& $0.16$\\\hline
Manipur& $0.34$\\\hline
Assam& $0.42$\\\hline
    \end{tabular}
\label{tab:India_variability}
\end{table}
\subsection{Variability of Temperature Change for Indian States and UTs}
Table~\ref{tab:India_variability} shows the variability of temperature change in the last decade for some of the Indian states and union territories (UTs). This variability is discussed in detail as follows:

%\subsubsection{States/UTs Most Affected by Temperature Change in the Last Decade} Lakshadweep suffered the most from temperature change in the last decade. Lakshadweep is a tropical archipelago (an extensive group of islands). Andaman and Nicobar Islands are other archipelagos on the list. It is already well known that islands are more vulnerable to and threatened by the effects of climate change. Tamil Nadu, Kerala, and Puducherry are all coastal regions in the southern part of the country. Climate change threatens coastal areas already stressed by human activity, pollution, invasive species, and storms. Himachal Pradesh, Uttarakhand, Delhi, Chandigarh, and Haryana are regions in northern India. North India has been witnessing frequent heatwaves and high temperatures recently.

%\subsubsection{States/UTs Least Affected by Temperature Change in the Last Decade} Tripura, Mizoram, Meghalaya, Manipur, and Assam form most of the northeast and are among the states that experienced the least temperature change in the country. West Bengal, Jharkhand, Orissa, and Bihar are eastern states of India, while Chhattisgarh lies in the central part.

\subsubsection{Overall Temperature Variability for Indian States \& UTs}
In the last decade (2011-2020), all the mentioned states \& UTs have had a positive temperature change, which means they witnessed an increase in temperature in the last decade. %Southern and northern regions have witnessed greater changes in temperature than the central or western parts, whereas Eastern and North-Eastern states have experienced lower changes.
It is already well known that islands are more vulnerable to and threatened by the effects of climate change.  Lakshadweep is a tropical archipelago (an extensive group of islands), hence witnessing the highest temperature change of 0.87$^{o}C$. Further, it can be observed that Tripura experienced the least change of 0.07$^{o}C$.  %Tamil Nadu, Kerala, and Puducherry are all coastal regions in the southern part of the country. Climate change threatens coastal areas already stressed by human activity, pollution, invasive species, and storms. Himachal Pradesh, Uttarakhand, Delhi, Chandigarh, and Haryana are regions in northern India. North India has been witnessing frequent heatwaves and high temperatures recently. 
Moreover, it is interesting to note that, on average, Indian states \& UTs have experienced a temperature change of more than 0.58$^{o}C$ over the last decade. %Out of these States \& UTs are seven southern, six northern, five western, two central, and two north-eastern states.

\subsubsection{Zone-Wise Temperature Variability Over Last Decade}
\paragraph{Highest Temperature Shifts-Southern Zone}
Tamil Nadu, Kerala, and Puducherry are all coastal regions in the southern part of the country. The southern zone encompasses coastal states, and islands like Lakshadweep and Andaman \& Nicobar Islands exhibited the most significant temperature changes. Its proximity to the coastline renders these regions more susceptible to the impact of climate change. The substantial rise in temperature in this zone underscores the vulnerability of coastal and island areas to environmental shifts.

\paragraph{Substantial Temperature Shifts-Northern Zone}
Northern India comprises of regions like Himachal Pradesh, Uttarakhand, Delhi, Chandigarh, and Haryana. Following closely, the northern zone registered noteworthy temperature changes, often experiencing recurrent heat waves and elevated temperatures. This trend indicates a consistent rise in temperatures within the northern regions, potentially signaling significant climate impacts.

\paragraph{Notable Temperature Alterations-Eastern and Central Regions}
West Bengal, Jharkhand, Orissa, and Bihar are eastern states of India, while Chhattisgarh lies in the central part.
The eastern and central parts of the country witnessed considerable temperature fluctuations exceeding 0.5$^{o}C$. These alterations signify notable shifts in temperature patterns within these regions, raising concerns about climatic changes in these areas.

\paragraph{Minimal Temperature Variations-North-Eastern States}
Conversely, the eastern and north-eastern states showed the least temperature changes over the period analyzed. Tripura, Mizoram, Meghalaya, Manipur, and Assam form most of the northeast and are among the states that experienced the least temperature change in the country. This region is characterized by its extensive lush forest cover and has exhibited relatively minor shifts in temperature. The significant forest cover, highlighted in the India State of Forests Report (ISFR) 2021, suggests a potential correlation between forested areas and limited temperature alterations in these states~\cite{data4}.

\subsection{Temperature Change Trend for Indian States and UTs}
Fig.~\ref{fig:wide_image} shows the trend of temperature change for Indian states and UTs. The observations are summarized as follows:

\paragraph{Continuous Temperature Rise}
Fig.~\ref{fig:wide_image} distinctly illustrates a persistent upward trend in temperature values from 1961 to 2020. A positive temperature change reflects a consistent increase in temperature levels over time.

\paragraph{Extremes in Temperature Fluctuations}
The years 2009 and 2010 stands out with the hottest winter and summer on record. This observation aligns with the findings of~\cite{climate3}, which suggested that 2009 and 2010 ranked among the top five years with the highest temperature in India. % the hottest year five years in which India had the highest temperature were 2009, 2010, 2016, 2017, and 2023
%The year 2009 stands out with the highest mean annual temperature change of 1.12$^{o}C$, indicating a notable spike in temperature during that year. In contrast, the year 1917 recorded the lowest temperature change of -0.88$^{o}C$, signifying a considerable decline in temperature during that period.

\paragraph{Erratic Temperature Patterns in Recent Years}
Notably, the graph portrays a relatively erratic and pronounced increase in temperature during the last two decades (2001-2020). This observed temperature rise suggests a period of more abrupt and irregular temperature fluctuations in the recent past.

\paragraph{Conclusive Temperature Trend}
Evaluating the comprehensive data from 1901 to 2020, it's evident that India has consistently experienced a progressive increase in temperature over this extensive period. The clear upward trajectory in temperature values reaffirms the country's ongoing trend of rising temperatures.

% \begin{figure}[h]
%   \centering
%   \includegraphics[width=\linewidth]{winter_t.png}
%   \caption{Trend Analysis for Summer and Winter in Delhi}
%   \label{fig4}
% \end{figure}
% \begin{figure}[h]
%   \centering
%   \includegraphics[width=\linewidth]{spring_t.png}
%   \caption{Trend Analysis for Spring and post monsoon for Delhi}
%   \label{fig5}
% \end{figure}
\begin{table}[]
  \centering
  \caption{Average Yearly Statistics for $T_{\text{max}}$ and $T_{\text{min}}$}
  %\resizebox{\columnwidth}{!}{%
    \begin{tabular}{|c|c|c|c|}
      \hline
      \textbf{Year} & \textbf{$T_{max}~(^{o}C)$} & \textbf{$T_{min}~(^{o}C)$} & \textbf{$\sqrt{
      \text{Var}(S)
      }~(^{o}C)$} \\
      \hline
      2016 & 47.05 & 4.05 & 6.15 \\
      2017 & 47.01 & 1.43 & 6.78 \\
      2018 & 46.48 & 2.01 & 6.63 \\
      2019 & 47.94 & 0.30 & 8.02 \\
      2020 & 47.08 & 1.64 & 6.98 \\
      \hline
    \end{tabular}%
  %}
  \label{tab:yearly_stats_tmax_tmin}
\end{table}

%\begin{table}[h]
%\centering
%\caption{Mann-Kendall Test Results for Seasons}
%\begin{tabular}{|c|c|c|c|c|}
%\hline
%\textbf{Season} & \textbf{Trend} & \textbf{h} & \textbf{p-value} & \textbf{Slope} \\ \hline
%Winter & Increasing & True & $1.57 \times 10^{-13}$ & $0.0014$ \\ \hline
%Summer & Increasing & True & $0.0$ & $0.0012$ \\ \hline
%post monsoon & Increasing & True & $0.0$ & $0.0048$ \\ \hline
%Spring & Decreasing & True & $6.63 \times 10^{-11}$ & $-0.0015$ \\ \hline
%\end{tabular}
%\vspace{5pt} % Adjust the spacing between the table and caption
%\label{tab:mann_kendall_seasons}
%\end{table}
\subsection{Variability and Temperature Change Trend for Delhi}
The yearly statistics for Delhi are shown in Table \ref{tab:yearly_stats_tmax_tmin}, where $T_{\text{max}}$, and $T_{\text{min}}$ are the maximum and minimum temperatures of each year. It can be observed that the minimum temperature decreases with each passing year. Further, the minimum temperature, $T_{min}$ decreases by $1.80^{o}C$ from 2016 to 2020. Moreover, the standard yearly deviation for 2019 is the highest, i.e., $8.02^{o}C$. 

The maximum and minimum temperature range from 2016-2020 is shown in Table \ref{tab:yearly_stats_tmax_tmin}. It can be observed that the average yearly temperature is highest for the year 2019, with $T_{max}= 47.94^{o}C$, and the average yearly minimum temperature is $0.30^{o}C$. Further, in Fig.~\ref{Seasonal_Trends_across_the_years}, the seasonal temperature trend for Delhi is shown. It can be observed that the mean temperature for Delhi has increased over the years. The climatic change for Delhi is erratic, with the summer season experiencing the highest temperature change of 3.1$^oC$ in 2010. This is in line with the fact that summers in 
2010 recorded the highest temperature change in India. %summers and winters has decreased by  1.4$^0C$ and 1.5$^0C$, respectively, from 1980 to 2022. This implies that summers (April-June) and winters (December-February) are becoming cooler over the years. However, the temperature for the monsoon season (July-Sepetember) has increased by 0.6$^0C$ over the years 1980-2022, and there has been no significant change in temperature for the post monsoon season (October-Novemeber) over the years.

\begin{figure}
  \centering
  \includegraphics[width=0.7\linewidth,trim={0 0 0 1.2cm},clip]{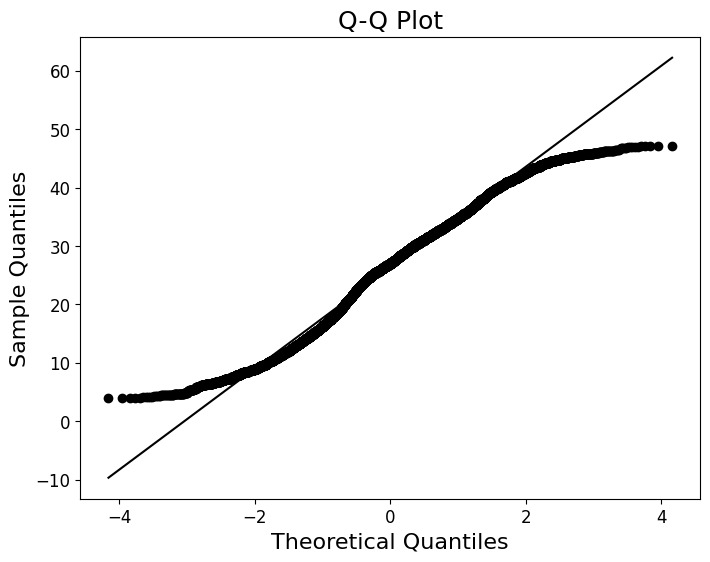}
  \caption{Q-Q plot for Delhi's temperature}
  \label{fig:q-qplot}
\end{figure}
\begin{table}
\centering
\caption{Mann-Kendall Test Result}
\begin{tabular}{|c|M{3 cm}|M{3 cm}|}
\hline
\textbf{Parameter} & \textbf{Significance}&\textbf{Value} \\
\hline
%Trend & &Increasing \\\hline
h & Hypothesis &True \\\hline
p & Compelling
evidence against the null hypothesis&0.000162 \\\hline
z & Strength of the evidence against
the null hypothesis&3.771 \\\hline
$\tau$&Movement in
the variable over the observed time frame & 0.0268\\\hline
%s &\textcolor{red}{....} & 1.030 ${\times 10^{6}}$ \\\hline
%Variance &-& 7.464 ${\times 10^{10}}$ \\\hline
%Slope &-& 6.978 ${\times 10^{-5}}$ \\\hline
%Intercept &-& 26.694 \\\hline
\end{tabular}
\label{tab:mann_kendall_test}
\end{table}
\subsection{Statistical Results}
Mann-Kendall and Theil Sen tests are more suited to non-normal distribution \cite{chattopadhyay2016long}. Thus, before applying the Mann-Kendall and Theil Sen tests for trend detection, we had to check for the normal distribution of data. %because non-parametric trend testing techniques were more suited to nonnormal distribution. 
To check for the normality of distribution, we utilized the p-value of Shapiro-Wilks test~\cite{test4}. Fig.~\ref{fig:q-qplot} plots a Q-Q plot, giving us a graphical visualization of normality. A straight line along the $x=y$ plane justifies the normal distribution of the data set. It can observed from Fig.~\ref{fig:q-qplot} that the plot is skewed at the ends, showing that the data is non-normal. The test results for the Mann-Kendall test are presented in Table~\ref{tab:mann_kendall_test}. The obtained p-value of 0.000162 stands below the conventional significance level of 0.05~\cite{test4}. This lower p-value signifies compelling evidence against the null hypothesis, providing substantial support for a significant trend within the dataset. The statistical significance, as indicated by the p-value, underscores that the observed trend is unlikely to result from random variability or chance alone. Moreover, the corresponding Z-value of 3.77 further corroborates the strength of the evidence against the null hypothesis. Higher Z-values indicate a more robust rejection of the null hypothesis, bolstering the assertion of a meaningful trend within the data. The calculated movement of the variable over the observed time frame, $\tau$ is 0.026. This complements these findings by signifying a positive correlation, reinforcing the notion of an increasing trend. This positive $\tau$ value denotes a consistent upward movement in the variable over the observed time frame. The slope of $6.978\times 10^{-5}$ and intercept of 26.694 further characterize this trend, indicating a steady and statistically significant increase over time. This suggests that the analyzed variable exhibits a consistent and noteworthy upward pattern over the observed period. The obtained p-value is significantly lower than the standard threshold, supporting the assertion that the observed trend is unlikely to be due to random fluctuations or chance alone. Therefore, it can be concluded that there is a genuine and substantial upward trend in the climatic changes across the years. %Moreover, the above statement can also be confirmed as indicated by the positive slope and confirmed by the statistical analysis. %The p-value for the Shapiro-Wilks test was higher than expected, rejecting the null hypothesis and proving that the data was not normally distributed. Fig.~\ref{Seasonal_Trends_across_the_years} has been used to plot a Q-Q plot. If the data had been normally distributed, the plot would have been , but . 

\section{Conclusions}
This paper provides a comprehensive analysis of India's climatic trends, emphasizing regional variations and specifically delving into the unique climate of Delhi. Specifically, this research unveils the temporal and spatial variations in temperature patterns by analyzing extensive datasets encompassing India's diverse landscapes. The study utilized advanced statistical tools and methodologies to scrutinize temperature's annual and seasonal variability. The regional analysis across India underscores substantial temperature shifts, with southern and northern areas experiencing significant changes while the eastern and north-eastern states exhibit minimal variations. Lakshadweep faced the highest change (0.87$^{o}C$), highlighting coastal vulnerability, while Tripura observed the least change (0.07$^{o}C$). Over the extensive period from 1901 to 2020, India consistently witnessed rising temperatures, particularly in recent decades, indicating irregular spikes and significant shifts, surpassing an increase of 0.58$^{o}C$ average over the years. Statistical analyses confirm a substantial and consistent upward trend in Delhi's climate variable, verified through Mann-Kendall test and related metrics, signifying a non-random increase over time. These findings are pivotal for policymakers, providing insights for tailored climate adaptation strategies, especially considering the vulnerability of coastal and island areas to climate change. Non-parametric tests like the Mann-Kendall test proved effective in discerning substantial trends within non-normally distributed data, emphasizing their significance in reliable trend detection amidst varied climatic dynamics. Future research should explore urbanization's impact on climate and monitor evolving climate patterns to facilitate informed decision-making and sustainable planning strategies.

\bibliography{main}

\begin{thebibliography}{10}
\providecommand{\url}[1]{#1}
\csname url@samestyle\endcsname
\providecommand{\newblock}{\relax}
\providecommand{\bibinfo}[2]{#2}
\providecommand{\BIBentrySTDinterwordspacing}{\spaceskip=0pt\relax}
\providecommand{\BIBentryALTinterwordstretchfactor}{4}
\providecommand{\BIBentryALTinterwordspacing}{\spaceskip=\fontdimen2\font plus
\BIBentryALTinterwordstretchfactor\fontdimen3\font minus \fontdimen4\font\relax}
\providecommand{\BIBforeignlanguage}[2]{{%
\expandafter\ifx\csname l@#1\endcsname\relax
\typeout{** WARNING: IEEEtran.bst: No hyphenation pattern has been}%
\typeout{** loaded for the language `#1'. Using the pattern for}%
\typeout{** the default language instead.}%
\else
\language=\csname l@#1\endcsname
\fi
#2}}
\providecommand{\BIBdecl}{\relax}
\BIBdecl

\bibitem{ref2}
\BIBentryALTinterwordspacing
A.~K. Gupta, M.~Negi, S.~Nandy, J.~M. Alatalo, V.~Singh, and R.~Pandey, ``Assessing the vulnerability of socio-environmental systems to climate change along an altitude gradient in the indian himalayas,'' \emph{Ecological Indicators}, vol. 106, p. 105512, 2019. [Online]. Available: \url{https://www.sciencedirect.com/science/article/pii/S1470160X19304972}
\BIBentrySTDinterwordspacing

\bibitem{climate_change_1}
G.~Wang, D.~Wang, K.~E. Trenberth, A.~Erfanian, M.~Yu, M.~G. Bosilovich, and D.~T. Parr, ``The peak structure and future changes of the relationships between extreme precipitation and temperature,'' \emph{Nature Climate Change}, vol.~7, p. 268–274, Apr 2017.

\bibitem{India1}
S.~Kreft, D.~Eckstein, and I.~Melchior, ``Spatiotemporal analysis of vegetation changes along the belt and road initiative region from 1982 to 2015,'' \emph{Global Climate Risk Index 2017 (PDF)}.

\bibitem{India2}
P.~R. Liu1 and A.~E. Raftery1, ``Country-based rate of emissions reductions should increase by 80\% beyond nationally determined contributions to meet the $2^{0}$ c target,'' \emph{PMC}, vol.~8, pp. 122\,579--122\,588, 2021.

\bibitem{India4}
``Warmer tibet can see brahmaputra flood assam | india news - times of india,'' Feb 2007.

\bibitem{India5}
``Climate change performance index,'' Nov 2022.

\bibitem{dash2009changes}
S.~Dash, M.~A. Kulkarni, U.~Mohanty, and K.~Prasad, ``Changes in the characteristics of rain events in india,'' \emph{Journal of Geophysical Research: Atmospheres}, vol. 114, no. D10, 2009.

\bibitem{singh2015trend}
D.~Singh, S.~K. Jain, and R.~D. Gupta, ``Trend in observed and projected maximum and minimum temperature over nw himalayan basin,'' \emph{Journal of Mountain science}, vol.~12, pp. 417--433, 2015.

\bibitem{shekhar2010climate}
M.~Shekhar, H.~Chand, S.~Kumar, K.~Srinivasan, and A.~Ganju, ``Climate-change studies in the western himalaya,'' \emph{Annals of Glaciology}, vol.~51, no.~54, pp. 105--112, 2010.

\bibitem{dad2021time}
J.~M. Dad, M.~Muslim, I.~Rashid, and Z.~A. Reshi, ``Time series analysis of climate variability and trends in kashmir himalaya,'' \emph{Ecological Indicators}, vol. 126, p. 107690, 2021.

\bibitem{ref3}
\BIBentryALTinterwordspacing
A.~Chakraborty, P.~Joshi, A.~Ghosh, and G.~Areendran, ``Assessing biome boundary shifts under climate change scenarios in india,'' \emph{Ecological Indicators}, vol.~34, pp. 536--547, 2013. [Online]. Available: \url{https://www.sciencedirect.com/science/article/pii/S1470160X13002446}
\BIBentrySTDinterwordspacing

\bibitem{climate2}
\BIBentryALTinterwordspacing
``The seasons, the equinox, and the solstices.'' [Online]. Available: \url{https://www.weather.gov/cle/Seasons}
\BIBentrySTDinterwordspacing

\bibitem{krishnanimpact}
R.~S. Krishnan, S.~Chaturvedi, and E.~Rajasekar, ``Impact of building design variables on natural ventilation potential and thermal performance: An evaluation in new delhi, india.''

\bibitem{data1}
\BIBentryALTinterwordspacing
``{NSRDB:} national solar radiation database.'' [Online]. Available: \url{https://nsrdb.nrel.gov/data-viewer}
\BIBentrySTDinterwordspacing

\bibitem{data2}
\BIBentryALTinterwordspacing
``Data catalog:climate knowledge portal.'' [Online]. Available: \url{https://climateknowledgeportal.worldbank.org/download-data}
\BIBentrySTDinterwordspacing

\bibitem{articleNN}
U.~Wadgave and M.~Khairnar, ``Parametric test for non-normally distributed continuous data: For and against,'' \emph{Electronic Physician}, vol.~11, pp. 2008--5842, 06 2019.

\bibitem{chattopadhyay2016long}
S.~Chattopadhyay and D.~R. Edwards, ``Long-term trend analysis of precipitation and air temperature for kentucky, united states,'' \emph{Climate}, vol.~4, no.~1, p.~10, 2016.

\bibitem{mann1945nonparametric}
H.~B. Mann, ``Nonparametric tests against trend,'' \emph{Econometrica: Journal of the econometric society}, pp. 245--259, 1945.

\bibitem{kendall1948rank}
M.~G. Kendall, ``Rank correlation methods.'' 1948.

\bibitem{MKandTS}
D.~Fan, L.~Ni, X.~Jiang, S.~Fang, H.~Wu, and X.~Zhang, ``Spatiotemporal analysis of vegetation changes along the belt and road initiative region from 1982 to 2015,'' \emph{IEEE Access}, vol.~8, pp. 122\,579--122\,588, 2020.

\bibitem{data3}
\BIBentryALTinterwordspacing
``Sign function.'' [Online]. Available: \url{https://en.wikipedia.org/wiki/Sign_function}
\BIBentrySTDinterwordspacing

\bibitem{data4}
\BIBentryALTinterwordspacing
``Forest survey report 2021.'' [Online]. Available: \url{https://pib.gov.in/PressReleasePage.aspx?PRID=1789635#:~:text=In%202021%2C%20the%20total%20forest,geographical%20area%20of%20the%20country.&text=Sharing%20the%20findings%2C%20the%20Minister,geographical%20area%20of%20the%20country.}
\BIBentrySTDinterwordspacing

\bibitem{climate3}
\BIBentryALTinterwordspacing
``Climate research \&services, pune.'' [Online]. Available: \url{https://www.imdpune.gov.in/lrfindex.php}
\BIBentrySTDinterwordspacing

\bibitem{test4}
\BIBentryALTinterwordspacing
J.~Jurečková and J.~Picek, ``Shapiro–wilk-type test of normality under nuisance regression and scale,'' \emph{Computational Statistics \& Data Analysis}, vol.~51, no.~10, pp. 5184--5191, 2007. [Online]. Available: \url{https://www.sciencedirect.com/science/article/pii/S0167947306002878}
\BIBentrySTDinterwordspacing

\end{thebibliography}
\bibliographystyle{IEEEtran}
\end{document}